# Evaluation of Applications Latency in Server Centric Passive Optical Network Based Data Centre Architectures

**Azza E. A. Eltraify, Mohamed O. I. Musa, Ahmed Al-Quzweeni and Jaafar M.H. Elmirghani**
*School of Electronic and Electrical Engineering, University of Leeds, Leeds, LS2 9JT, UK*
*E-mail: {scaeae, m.musa, ml13anaq, j.m.h.elmirghani}@leeds.ac.uk*

**ABSTRACT**

The number of applications running in the cloud has dramatically increased in the past decade as well as the number of users accessing them. Data centres resources, architectures and conditions define the performance of the applications running on them. One of the main measures of the network efficiency is the latency, which can have a huge effect on resources utilisation, power consumption and the overall performance. In this paper, the performance of a fog network is evaluated by measuring the latency while running a facial recognition software. The network consists of two processing cells, a core network and a PON cell. The results show how network latency is affected by running the facial recognition software in the end-to-end network setup introduced in this paper.

## 1. INTRODUCTION

Data centre architectures have been developed and optimized throughout the year to meet the market demand of growing data centres. Research and studies have been conducted to ensure that data centre architectures are efficient, scalable and more resilient to adapt to the increasing demands [1]-[10]. A key step in the deployment of these architectures is the experimental evaluation of these designs.

Passive optical networks (PON) architectures have been a step towards more efficient networks providing lower power consumption, higher speed and data rates, hence providing lower latency. The architecture demonstrated in this paper is a proposed high-speed server centric architecture as shown in figure 1 [11]. The architecture is composed of several PON cells, each cell contains several racks of servers communicating through an OLT. Communication among servers of different groups is provisioned through an optical backplane [12]. For intra-rack communication, traffic aggregation is carried out by designated relay servers in each group, while communication between the OLT and the racks is provisioned through a coupler (Time Division Multiplexing (TDM)) or via an arrayed Waveguide Grating (AWGR) through Wavelength Division Multiplexing (WDM) [11], [13] - [15].

One of the main issues facing data centre architectures is maintaining low latency which greatly impacts the performance of applications running on them. Applications differ in requirements for resources. Applications that require real time video streaming are considered some of the most demanding applications that exhaust the resources of networks. The application chosen for this paper is a facial recognition software. Facial recognition software operates by going through a video stream to detect faces, capturing them, then analysing and comparing patterns to data either stored in a cloud or a local database. Once a face is captures the analog information is transformed into digital information then sent over for pattern analysis. Facial recognition is considered a very important application in several areas, such as security, law enforcement, health, biometrics, marketing and retail. Thales has been the leading company in biometrics analysis and facial recognition technology. Thales' Live Face Identification System (LFIS) is considered the most advanced solution in the market with face acquisition rate of 99.44% in less than 5 seconds [16].

In this paper an experimental setup of an end-to-end network of over 110 km of optical interconnections is setup while running a facial recognition software on one end and a live stream from an IPTV camera at the other end. The experiment evaluates the effects of facial recognition software on the latency of the network. The results are compared to measurements without the application running on the experiment.

## 2. EXPERIMENTAL SETUP

The experimental setup consisted of two processing cell, core network, PON cell, and the setup of facial recognition software. The processing cells are based on a server centric passive optical network data centre architecture as shown in Figure 1. The architecture consisted of three racks each with three servers, the communication among racks was achieved using a 10 Gbps Cisco Switch and media converters as an alternative to the optical polymer backplane [12]. The implementation of the architecture and the addressing is shown in Figure 2. The routing functionalities were implemented in servers using a specialised routing operating system called MikroTik OS.

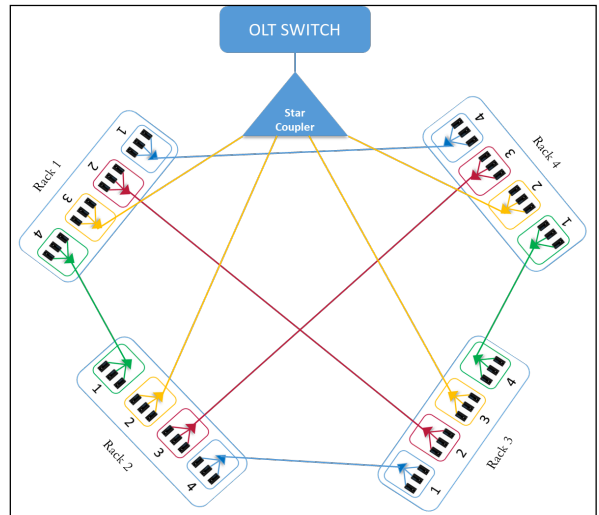

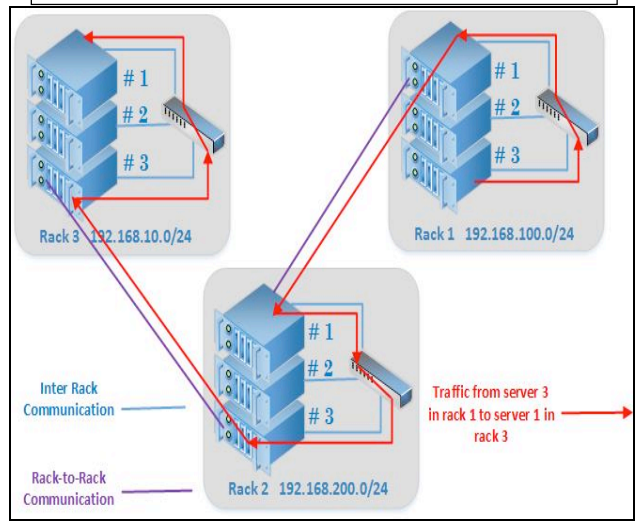

*Figure 1. Server Centric PON based data centre network architecture [13]*

*Figure 2. Intra-rack and inter-rack communication in the proposed PON based data centre architecture*

The core network implementation was based on an IP over DWDM core network of three nodes connected via over 100 km of optical fibre links as shown in Figure 4 [18]-[29]. The nodes are equipped with 100 Gbps MRV/ADVA DWDM nodes. The nodes simulate long distances between cities, which provides the ability to run actual tests representing different setups of data centres [30]-[38].

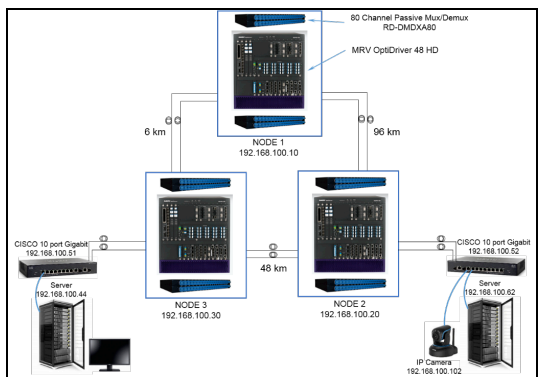
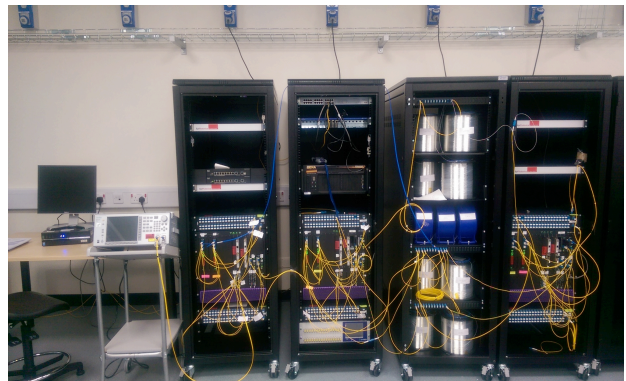

*Figure 4. IP/WDM Core Network Nodes*

Figure 5. shows the end-to-end communication of the proposed setup with two PON processing cells one at each end, communicating via an IP/WDM core network. The live video stream was captured through an IPTV

Camera connected at one end and the facial recognition software was setup on a computer connected to the other end.

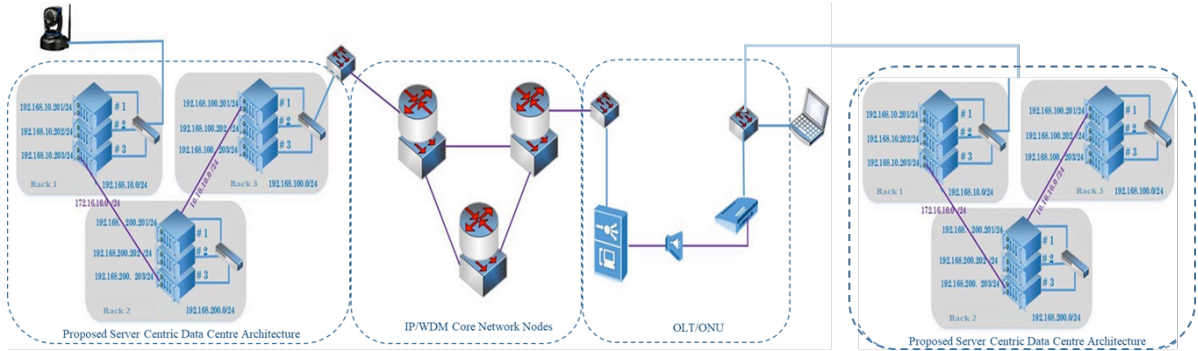

*Figure 5. The implemented end-to-end Server Centric Data Centre Architecture, IP/WDM Core Network Nodes and OLT/ONU*

### 3. RESULTS

In this paper, we study the impact of running a facial recognition software on network latency in an end-to-end network architecture setup composed of two processing cells, IP/WDM Core network nodes and a PON cell. The results in this paper demonstrate the difference in latency in the experimental setup with and without the facial recognition software. It demonstrates the trade-offs between network latency and applications. Table 1. shows the typical latency in conventional mediums of communication that are used in this experiment. The real time video feed was captured through an IPTV camera, and received at the other end by the facial recognition software over a distance of 110 km.

| Network Element | Measurement | Typical Latency |
| --- | --- | --- |
| ToR L2 Switch | 10G / 100G | 300ns-500ns |
| Propagation Delay | 100m Fiber | 490ns |
| Low Latency NIC | 10G | 1μs |
| Serialization Delay | 1518B over 10G link | 1.25μs |

*Table 1. Typical Latency of Network Equipment and Mediums.*

The latency in the architecture was measured using ICMP packets type 0 (ping request). The signal was sent from the node connected to the IPTV camera to the node with the facial recognition software. The latency measurements at each node were captured by measuring the round-trip time (RTT). The test was conducted 10 times to test the effect of different traffic loads on the latency. Figure 6. shows the latency in the experimental setup without running the facial recognition software [40], while Figure 7. shows the latency in the architecture after the deployment of the facial recognition software.

Comparing Figure 6. and Figure 7. there is a noticeable increase in latency of ~ 85%. Despite the high increase in latency, the architecture and setup have proven to be efficient with the average highest latency being 3.5 ms and the lowest latency was 0.39ms, which is quite low compared to conventional data centre networks architectures setups.

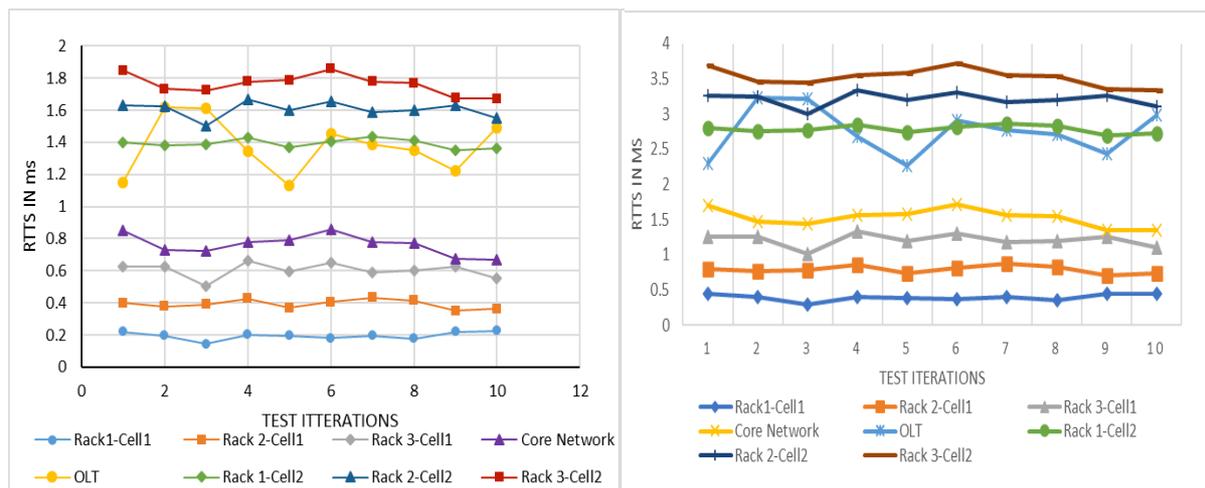

*Figure 6. The round-trip time of all node (No software)*     *Figure 7. The round-trip time of all node (Software)*

## 4. CONCLUSIONS

This paper demonstrated an experimental setup to evaluate the ramifications of applications on network latency over server centric passive optical network based data centre architecture. The application used in this experiment was a facial recognition software running at one end of the network while receiving a live video stream from an IPTV camera at the other end. The latency was measured using ICMP signals to capture the round time trip of packets. The results have shown increase in latency after the deployment of the facial recognition software. The maximum latency captured was less than the latency in conventional data centre architectures, which shows that the proposed architecture is more efficient.

ACKNOWLEDGEMENTS

The authors would like to acknowledge funding from the Engineering and Physical Sciences Research Council (EPSRC), INTERNET (EP/H040536/1) and STAR (EP/K016873/1) and from the EPSRC TOWS (EP/S016570/1) project. All data are provided in full in the results section of this paper.